\documentclass[acmsmall,screen,nonacm]{acmart}

\usepackage{xspace,cleveref}
\usepackage{algorithm}
\usepackage{algpseudocode}
\usepackage{listings}

\lstdefinestyle{paperlisting}{
  basicstyle=\small\ttfamily,
  breaklines=true,
  columns=fullflexible,
  frame=single,
  keepspaces=true,
  showstringspaces=false,
  xleftmargin=1em,
}

\begin{document}

\author{Yumeng He}
\orcid{0009-0002-1171-4191}
\email{u1528477@umail.utah.edu}
\affiliation{%
  \institution{University of Utah}
  \city{Salt Lake City}
  \state{UT}
  \country{USA}
}

\author{Pavel Panchekha}
\orcid{0000-0003-2621-3592}
\email{pavpan@cs.utah.edu}
\affiliation{%
  \institution{University of Utah}
  \city{Salt Lake City}
  \state{UT}
  \country{USA}
}

\keywords{Floating point, debugging, double-double, numeric, re-execute}

\newcommand{\todo}[1]{{\color{orange}[#1]}\xspace}
\newcommand{\yumeng}[1]{{\color{blue}[#1]}\xspace}
\newcommand{\pavel}[1]{{\color{red}[#1]}\xspace}
\DeclareRobustCommand{\name}{RePo\xspace}
\DeclareRobustCommand{\RO}{residue override\xspace}
\DeclareRobustCommand{\PRO}{parallel residue override\xspace}
\DeclareRobustCommand{\eftsan}{EFTSanitizer\xspace}
\DeclareRobustCommand{\eftsanb}{EFTSanitizer-Buggy\xspace}
\DeclareRobustCommand{\eftsanf}{EFTSanitizer-Fixed\xspace}
\newcommand{\NumTotalBenchmarksWithNonFP}{56\xspace}
\newcommand{\numNASBenchWithNonFP}{8\xspace}
\newcommand{\numRodiniaBenchWithNonFP}{18\xspace}
\newcommand{\numPolybenchBenchWithNonFP}{30\xspace}

\newcommand{\numNASBenchNonFP}{0\xspace}
\newcommand{\numRodiniaBenchNonFP}{5\xspace}
\newcommand{\numPolybenchBenchNonFP}{2\xspace}

\newcommand{\numTimeoutBenchmarks}{5\xspace} 

\newcommand{\numTotalBenchmarks}{44\xspace}
\newcommand{\numNASBench}{6\xspace}
\newcommand{\numPolybenchBench}{28\xspace}
\newcommand{\numRodiniaBench}{10\xspace}

\newcommand{\numSciSolvedBenchmarks}{30\xspace}
\newcommand{\numSciInterestingBenchmarks}{14\xspace}
\newcommand{\numSciBenchmarksImproved}{13\xspace}
\newcommand{\numSciBenchmarksMixed}{1\xspace}
\newcommand{\numSciBenchmarksGreat}{3\xspace}
\newcommand{\numSciBenchmarksPerfect}{10\xspace}
\newcommand{\pctSciBenchmarksPerfect}{71.43\%\xspace}

\newcommand{\numBenchmarksTriggerRO}{8\xspace}

\newcommand{\avgReexecutions}{3.6\xspace}
\newcommand{\capReexecutions}{20\xspace}

\newcommand{\numMicrobench}{45\xspace}

\newcommand{\numFPBenchInteresting}{42\xspace}
\newcommand{\numFPBenchImproved}{32\xspace}
\newcommand{\numFPBenchPerfect}{20\xspace}
\newcommand{\numFPBenchMixed}{6\xspace}
\newcommand{\numFPBenchLoss}{4\xspace}

\newcommand{\numMFBenchmarks}{169\xspace}
\newcommand{\numMFInteresting}{34\xspace}
\newcommand{\numMFNonRO}{106\xspace}
\newcommand{\numROInteresting}{29\xspace}
\newcommand{\numROTotalImproved}{25\xspace}
\newcommand{\pctROTotalImproved}{86.2\%\xspace}
\newcommand{\numROStrictImproved}{19\xspace}
\newcommand{\numROPerfect}{8\xspace}
\newcommand{\numROMixed}{6\xspace}
\newcommand{\numRONoHelp}{4\xspace}

\newcommand{\numFPBenchNonRO}{20\xspace}
\newcommand{\numFPBenchNonROImproved}{16\xspace}
\newcommand{\numFPBenchNonROLoss}{4\xspace}
\newcommand{\numFPBenchROWin}{16\xspace}
\newcommand{\numFPBenchROPerfect}{5\xspace}
\newcommand{\numFPBenchROMixed}{6\xspace}

\newcommand{\numQDInteresting}{57\xspace}
\newcommand{\numQDImproved}{43\xspace}
\newcommand{\pctQDImproved}{75.44\%\xspace}
\newcommand{\numQDMixed}{6\xspace}
\newcommand{\numQDLoss}{8\xspace}

\title{Accurate Residues for Floating-Point Debugging}

\begin{abstract}
Floating-point arithmetic is error-prone and unintuitive.
Floating-point debuggers instrument programs to
monitor floating-point arithmetic at run time
and flag numerical issues.
To do so, they estimate \emph{residues}---%
the difference between
actual floating-point and ideal real values---%
for every floating-point value in the program.
A large literature has explored various approaches
for computing these residues accurately
(leading to few false reports, i.e.,
false positives and false negatives)
and efficiently (leading to low overhead over uninstrumented execution).
Unfortunately, the most efficient methods,
based on ``error-free transformations'',
have a high rate of false positives,
while the most accurate methods,
based on high-precision arithmetic,
are very slow.
This paper builds on error-free-transformations-based approaches
and aims to improve their accuracy while preserving efficiency.

To more accurately compute residues,
this paper divides residue computation into two steps---%
rounding error computation and residue function evaluation---%
and shows how to perform each step accurately
via careful improvements to the current state of the art.
We evaluate on \numTotalBenchmarks~large scientific computing workloads,
focusing on the \numSciInterestingBenchmarks~benchmarks
where prior tools produce false reports:
our approach eliminates false reports on \numSciBenchmarksPerfect~benchmarks
and substantially reduces them on the remaining \numSciBenchmarksGreat~benchmarks.

Moreover, we find that
more complex numerical issues,
such as those found in numerical analysis textbooks,
require additional care,
because floating-point debuggers suffer from \emph{absorption},
in which two different machine-precision residues
cannot both be computed accurately in a single execution.
To address absorption, this paper introduces \emph{\RO},
which re-executes the program multiple times,
computing different residues in different executions
and assembling a final ``patchwork'' execution
where all residues are accurately computed.
We evaluate on \numMFBenchmarks~standard benchmarks
drawn from numerical analysis papers and textbooks,
requiring only \avgReexecutions~re-executions on average.
Among \numMFInteresting~benchmarks with false reports in the initial run,
\RO is triggered on \numROInteresting~of them
and reduces false reports on \numROTotalImproved~of them,
averaging 7.1 re-executions.
\end{abstract}

\maketitle
\section{Introduction}

Scientific and mathematical software
  typically uses floating-point arithmetic
  to approximate arithmetic on real numbers.
However, floating-point arithmetic is
  highly unintuitive~\cite{handbook-fp}.
Numerical problems like cancellation
  can invalidate whole computations,
  leading to potentially catastrophic consequences:
  financial losses~\cite{distort-stock,wall-street-distort-stock},
  election instability~\cite{election-fp},
  and wartime casualties~\cite{patriot-fp}.
The programming languages community has therefore
  long worked on tools, such as floating-point debuggers,
  that make numerical programming safer and easier.

Floating-point debuggers instrument a numerical program
  to detect, at run time, inaccurate operations that cause numerical issues.
A variety of floating-point debugging approaches
  have been developed over the years,
  including high-precision shadow execution~%
  \cite{fpdebug,herbgrind,fpsanitizer,pfpsanitizer},
  error-free transformations~\cite{eftsan,explanifloat,atomu,baozhang,shaman},
  and stochastic rounding~\cite{verrou}.
At their core, these debuggers aim to estimate \emph{residues},
  meaning the difference between the value of a variable
  in the actual floating-point execution of a program
  and in an idealized real-number execution.
By estimating these residues, floating-point debuggers can
  warn about values with large residues,
  detect unstable control flow,
  or track how erroneous operations affect program outputs.

The challenge is estimating residues accurately and efficiently.
Floating-point debuggers using high-precision arithmetic,
  such as FpDebug~\cite{fpdebug}, Herbgrind~\cite{herbgrind},
  and FPSanitizer~\cite{fpsanitizer,pfpsanitizer},
  are highly accurate but cause overheads of $100\times$ or more
  over uninstrumented execution.
Debuggers based on error-free transformations
  use only machine floating-point operations
  and are much faster~\cite{eftsan,explanifloat,atomu,baozhang},
  but suffer a higher rate of false positives and negatives,
  with as many as 20\% of warnings being false~\cite{explanifloat}.
Stochastic debuggers require repeated re-executions
  to achieve accurate results~\cite{verrou}.
To date, no floating-point debugging technique
  simultaneously achieves both high accuracy
  and low runtime overhead.

This paper proposes \name,
  and demonstrates that this best of both worlds
  is achievable with this approach to floating-point debugging.
\name splits residue computation into two steps---%
  \emph{rounding error estimation} using error-free transformations
  and \emph{residue function evaluation} using machine floating-point operations---%
  and carefully addresses accuracy challenges for each step,
  such as accurate rounding errors for casts and rounding functions
  and accurate residue functions for multiplication operations.
Compared to the current state of the art,
  which produces false reports on \numSciInterestingBenchmarks~of
  \numTotalBenchmarks~scientific benchmarks,
  these improvements eliminate false reports on \numSciBenchmarksPerfect~benchmarks
  and substantially reduce them on the remaining \numSciBenchmarksGreat~benchmarks.

While accurate rounding errors and residue functions
  significantly improve debugging accuracy,
  addressing the most challenging numerical issues
  requires additional techniques.
Specifically, machine-precision residues,
  which are critical for high performance,
  suffer from \emph{absorption},
  meaning that, for some programs, two different residues
  cannot both be accurately measured in the same execution.
To address absorption, \name introduces \RO,
  a technique that performs \emph{multiple} executions of the program,
  computing a \emph{different} set of residues accurately each time.
A final execution then combines residues,
  measured in different executions,
  to produce an accurate floating-point debugging result.
Specifically, \name uses a carefully designed algorithm
  to resolve many instances of absorption in a small number of re-executions,
  even for complex numerical computations
  such as the internals of elementary functions.
Across \numMFInteresting~of \numMFBenchmarks~standard benchmarks
  drawn from numerical analysis papers and textbooks,
  where \name produces false reports in the initial run,
  \RO is triggered on \numROInteresting~of them
  and reduces false reports on \numROTotalImproved~of them,
  averaging 7.1 re-executions.

\medskip
\noindent
In short, this paper's contributions are:
\begin{enumerate}
\item Demonstrating that
  careful rounding error estimation and residue function evaluation
  significantly reduce false positive and false negative rates
  on large-scale scientific software (\Cref{sec:ops}).
\item Demonstrating that, despite this,
  absorption makes it impossible to avoid
  false positives and false negatives
  on complex numerical benchmarks (\Cref{sec:override}).
\item Demonstrating the \RO technique,
  which computes residues accurately,
  even in the presence of absorption,
  through multiple re-executions (\Cref{sec:impl} and \Cref{sec:pro}).
\end{enumerate}      
\section{Worked Example}
\label{sec:overview}

Consider the following computation
  over a double-precision variable \texttt{x}:
\begin{verbatim}
double a = x + 1;
double b = sqrt(x), c = sqrt(a);
double y = c - b;
double z = y * y;
\end{verbatim}

For an extreme input like $x=10^{99}$
  this code outputs $z = 0$,
  whereas the true value of $z$ is approximately $2.5\cdot10^{-100}$.
A floating-point debugger should help the user diagnose this issue.
  
\paragraph{Residues}

\begin{table}
  \caption{
    A step-by-step demonstration
      of how an ideal floating-point debugger should work
      for the running example.
    Each operation has an actual value,
      an ideal value,
      and a residue, which is the difference between the two.
    In this case, the residues for \texttt{b} and \texttt{c}
      are both large and of similar magnitude.
    The debugger warns the user about values
      (here, \texttt{y} and \texttt{z})
      whose residues are relatively large
      compared to their actual values.
  }
  \label{tbl:mpfr}
  \centering
  \begin{tabular}{lrrr}
    \hline
    Operation & Actual value & Ideal value & Residue \\
    \hline
    \texttt{a=x+1} &
    1.00000000e+99 &
    1.00000000\ldots00100\ldots{}e+99 &
    1 \\
    \hline
    \texttt{b=sqrt(x)} & 
    3.16227766e+49 &
    3.16227766\ldots29518\ldots{}e+49 &
    1.31447527\ldots18810\ldots{}e+32 \\
    \hline
    \texttt{c=sqrt(a)} & 
    3.16227766e+49 &
    3.16227766\ldots29676\ldots{}e+49 &
    1.31447527\ldots18968\ldots{}e+32 \\
    \hline
    \texttt{y=c-b} & 
    0 &
    1.58113883\ldots14719\ldots{}e\textminus50 &
    1.58113883\ldots14719\ldots{}e\textminus50 \\
    \hline
    \texttt{z=y*y} & 
    0 &
    0.24999999\ldots99875\ldots{}e\textminus99 &
    0.24999999\ldots99875\ldots{}e\textminus99 \\
    \hline
  \end{tabular}
\end{table}

Floating-point debuggers work by computing the \emph{residue}
  for each floating-point value in the program.
Formally, the residue is the difference between
  two executions of the program:
  the actual value computed in a floating-point execution of the program,
  and the ideal value computed in a hypothetical real-number execution.
Every floating-point operation in the program has an associated residue;
  \Cref{tbl:mpfr} shows the residue value for each operation
  in our running example.
For example, for the addition $a = x + 1$,
  the actual floating-point result is $10^{99}$
  while the ideal real-number result is $10^{99} + 1$,
  so the residue is $1$.
Floating-point debuggers use these residues
  to warn users about potential numerical issues.
For example, since the residue for $z$,
  roughly $2.5\cdot10^{-100}$,
  is large relative to its computed value of $0$,
  a floating-point debugger can issue a warning
  that the value of $z$ is unreliable and subject to numerical error.
The exact threshold used to trigger warnings
  varies across debuggers~\cite{fpsanitizer,herbgrind},
  but accurate residue computation is essential for all of them.

Various techniques for estimating residues exist.
Early floating-point debuggers like Herbgrind~\cite{herbgrind}
  attempted to directly compute the ideal value using MPFR~\cite{mpfr},
  and then subtract it from the actual value.
This required hundreds or even thousands of bits of precision
  to obtain accurate results.
Other debuggers, such as EFTSanitizer~\cite{eftsan},
  attempt to compute residues directly
  using \emph{error-free transformations},
  which are clever sequences of machine floating-point operations
  that capture the error of other machine floating-point operations.
This requires less precision and is thus much faster,
  but the resulting residues are less accurate,
  causing many false positives and false negatives compared to the MPFR-based approach.
There are other approaches as well;
  for example, Verrou~\cite{verrou} estimates residues
  by aggregating many randomized executions of the program.
In contrast, this paper separates residue computation
  into two distinct tasks:
  accurate estimation of \emph{rounding errors},
  at which error-free transformations excel,
  and their aggregation via \emph{residue functions}.

\paragraph{Rounding Errors and Residue Functions}
The residue $e_z$ of a floating-point operation
  $\hat{z} = \hat{f}(\hat{x}, \hat{y})$
  comes from two effects.
First, the real value $f(\hat{x}, \hat{y})$
  might not be exactly representable in floating-point,
  so $f(\hat{x}, \hat{y})$ must be rounded.
That rounding introduces the \emph{rounding error}
  $\mu_z = f(\hat{x}, \hat{y}) - \hat{f}(\hat{x}, \hat{y})$.
Second, since $\hat{x}$ and $\hat{y}$
  are themselves floating-point values,
  they are affected by their own rounding errors,
  and recursively by even earlier rounding errors
  of earlier intermediate values.
Conceptually,
  the residue $e_z$ aggregates rounding errors
  as $e_z[\mu_1, \ldots, \mu_x, \mu_y, \mu_z]$.
It is computed via a residue function
  $e_f(\hat{x}, \hat{y}, \hat{z}, \mu_z, e_x, e_y)$
  together with the original floating-point operation.
The process of computing rounding errors is relatively straightforward,
  using \emph{error-free transformations}
  that estimate rounding errors $\mu_z$
  for elementary operations such as addition/subtraction,
  multiplication, division, and square root.
Residue functions, however, are more challenging.

Previous floating-point debuggers
  that use error-free transformations,
  such as EFTSanitizer and Shaman~\cite{eftsan,shaman},
  focus on simply detecting
  whether a program is accurate or not.
This is a coarser objective than
  accurately computing all residues in the program,
  and both tools thus use correspondingly coarse,
  \emph{first-order} residue functions.
For example, consider the squaring operation,
  $\hat{z} = \hat{y}\ \hat{\cdot}\ \hat{y}$,
  in our example program.
The residue $e_z$ is computed as
\[
  z - \hat{z}
  = y \cdot y - \hat{y}\ \hat{\cdot}\ \hat{y}
  = (\hat{y} + e_y) \cdot (\hat{y} + e_y) - (\hat{y} \cdot \hat{y} - \mu_z)
  = \mu_z + 2 \hat{y} e_y + e_y^2,
\]
  but a first-order residue function drops
  the $e_y^2$ term.
While these first-order residue functions
  typically compute \emph{some} residues accurately,
  they can lead to significant inaccuracies
  as errors interact and compound.
For example, in the running example,
  the actual value of $y$ is zero
  even though its residue $e_y$ is nonzero.
With the first-order residue function,
  $e_z$ is therefore computed to be zero,
  whereas an accurate residue function yields a nonzero value.
More generally,
  we find that computing residue functions accurately
  significantly reduces the number of
  false positive and false negative warnings
  on a suite of standard scientific computing benchmarks.

\paragraph{Absorption}

\begin{table}
  \caption{
    A step-by-step demonstration of how \name uses \RO to resolve absorption.
    The first run produces two false negatives at the operations computing
      \texttt{y} and \texttt{z}
      because absorption occurred when computing $e_c$.
    The \RO mechanism therefore detects the operations whose rounding errors
      have the largest impact on the cancelled residues---in this case,
      the two square-root operations---and silences them in the next
      execution (Run~2).
    As a result, the residue for $\sqrt{x}$ becomes $0$,
      while the residue for $\sqrt{x+1}$ becomes the previously absorbed
      small term (about $1.6\times10^{-50}$).
    At this run, although we correctly measure the residues for
      the last two operations (which were previously false negatives),
      the residues at the two square-root operations become much worse.
    Therefore, in the final execution (Run~3),
      \RO stops silencing the two square-root operations
      and overrides the residues of the last two operations
      with the previously correctly computed values.
    Now every operation has an accurate residue.
  }
  \label{tbl:ro}
  \centering
  \begin{tabular}{lrrrr}
    \hline
    Operation & Residue (Run 1) & Residue (Run 2) & Residue (Run 3)\\
    \hline
    \texttt{a=x+1} &
    1.0000000000000000e+00 &
    1.0000000000000000e+00 &
    1.0000000000000000e+00 \\ \hline

    \texttt{b=sqrt(x)} & 
    1.3144752779492117e+32 &
    0 &
    1.3144752779492117e+32 \\ \hline

    \texttt{c=sqrt(a)} & 
    1.3144752779492117e+32 &
    1.5811388300841897e\textminus50 &
    1.3144752779492117e+32 \\ \hline

    \texttt{y=c-b} & 
    0 &
    1.5811388300841897e\textminus50 &
    1.5811388300841897e\textminus50 \\ \hline

    \texttt{z=y*y} & 
    0 &
    0.2500000000000000e\textminus99 &
    0.2500000000000000e\textminus99 \\ \hline
  \end{tabular}
\end{table}

Even with maximally accurate
  rounding errors and residue functions,
  some inaccuracies remain
  in the most challenging programs,
  because of a phenomenon we name \emph{absorption}.
Consider two different residues in our running example:
  $e_c$ and $e_y$,
  which are the residues of operations computing
  $c = \sqrt{a}$ where $a = x+1$, and $y = c-b$ where $b=\sqrt{x}$.
The residue $e_c$ is affected by
  the rounding error $\mu_c$ of the square root operation
  and the rounding error $\mu_a$ of the addition operation.
Moreover, the effect of $\mu_c$ is much, much larger
  than the effect of $\mu_a$—by about 50 orders of magnitude!
Accurately measuring the residue $e_c$ thus requires
  accurately computing $\mu_c$
  and accurately incorporating it with $\mu_a$ into $e_c$.

However, since $\mu_c$ is \emph{so} much larger than the
  propagated contribution of $\mu_a$,
  this forces $e_c$ to round away the contribution of
  $\mu_a / 2c$ entirely.
We say that $\mu_c$ \emph{absorbs} this contribution;
  it is an unavoidable consequence of storing the residue $e_c$
  in finite machine precision while attempting to compute it accurately.
The problem comes later, when attempting
  to compute $e_y = e_c - e_b$.%
\footnote{The original formula should be $e_y = \mu_y + e_c - e_b$
  but we elide $\mu_y$ because it is equal to $0$.}
The ideal result is roughly $\mu_a / 2c$,
  but because $\mu_c$ entirely absorbed this contribution
  when computing $e_c$,
  the actual computed residue $e_y$ becomes $0$.
In other words, attempting to compute $e_c$ accurately
  results in $e_y$ being computed inaccurately.
Alternatively, it is possible to compute $e_y$ accurately
  by simply ignoring the contribution of $\mu_c$ to $e_c$
  and $\mu_b$ to $e_b$.
Then $e_c$ will be computed to be $\mu_a / 2c$,
  while $e_b$ will be computed to be $0$,
  since we effectively ignore all contributions to it.
This results in highly inaccurate residues for $e_c$ and $e_b$,
  but it \emph{does} result in an accurate value for $e_y$:
  $\mu_a / 2c - 0 = \mu_a / 2c$.
In short, absorption means
  that either $e_c$ or $e_y$ can be computed accurately,
  but not both in the same execution.

\paragraph{Residue Override}

The impossibility of accurately measuring
  both $e_c$ and $e_y$ in the same execution
  in error-free-transformations-based debuggers,
  due to the problem of absorption,
  suggests an alternative approach:
  measure each accurately in separate executions
  and then combine the results.
\name's \emph{residue override} technique does exactly this.
During the first execution,
  it \emph{detects} residues
  that cannot be accurately computed in the same execution.
During a second execution,
  it \emph{silences} the rounding errors
  with the greatest influence on one set of residues
  by setting those rounding errors to $0$,
  and \emph{probes}, meaning it records,
  the now accurately computable residue.
Finally, in a third execution,
  the rounding errors are no longer silenced,
  but the inaccurate residues are \emph{overridden}
  with the more accurate probed values.
This final execution has accurate residues for each value
  and thus produces fewer false positives and false negatives.
\Cref{tbl:ro} shows how \RO in \name resolves the problem of absorption
  and produces correct residues
  for all operations in the running example.
Our results, described in \Cref{sec:eval},
  show that \name works remarkably well
  even on challenging programs,
  greatly reducing the number of
  false positives and false negatives
  on a standard suite of benchmarks
  drawn from textbooks and numerical analysis papers.

\section{Accurate Machine-Float Residues}
\label{sec:ops}

Floating-point debuggers rely on accurate residues to determine
  when to warn users about numerical issues.
To accurately compute residues for all floating-point intermediate values,
  \name aims to estimate rounding errors for each floating-point operation
  and then use those rounding errors to evaluate the residue function
  for each operation.
\name's implementation of these two steps is based on \eftsan,
  and the overall approach is similar.
However, our initial evaluation showed that \eftsan
  commonly produced false positives and false negatives.
By refining rounding error estimation and residue function evaluation,
  \name produces dramatically more accurate residues,
  with few false positives and false negatives
  across a range of scientific software.

\subsection{Accurate Rounding Error Estimation}

Rounding error estimation means computing,
  for an operation $f(\hat{x}, \hat{y})$ on floating-point inputs $\hat{x}$ and $\hat{y}$,
  the difference between
  the exact real result and the actual floating-point result.
In other words, it estimates the error introduced
  \emph{by that operation}.
We refer to this as rounding error \emph{estimation}
  because for some operations, such as division and square root,
  the rounding error is not exactly representable in floating-point,
  and is therefore itself rounded.
For the basic arithmetic operations---%
  addition, subtraction, multiplication, division, and square root---%
  \name estimates rounding error identically to \eftsan,
  using error-free transformations.

However, \name also estimates rounding error
  for casts between data types.
When a 64-bit floating-point value
  is cast to a 32-bit floating-point value, for example,
  the resulting 32-bit value
  may not be exactly the same as the original 64-bit value,
  since fewer bits are available,
  and thus rounding error occurs.
\name instruments all such cast operations
  and measures error by casting the 32-bit result
  back to 64 bits and subtracting.
The subtraction is exact by Sterbenz's law.
Casts from 32 to 64 bits are exact
  and do not need to be instrumented.

Additionally, \name detects and specially handles
  a numerical trick for rounding double-precision values.
In standard round-to-nearest-even rounding,
  when $|x| < 2^{52}$,
  adding and then subtracting the value $c = 1.5\cdot2^{52}$,
  as in $(x + c) - c$,
  rounds the value $x$ to its nearest integer.
This works because adding $c$ rounds off
  all fractional bits of $x$.
Since $c$ is itself an integer,
  subtracting it restores the value of $x$
  with the fractional bits rounded away.
Importantly, since the purpose of this trick
  is to round $x$ to an integer,
  the rounding errors of the individual
  addition and subtraction operations
  are irrelevant: they are purposeful,
  not accidental,
  and measure the fractional bits of $x$,
  which are intentionally discarded.
\name detects this specific pattern
  (as well as the equivalent single-precision pattern,
   and variants in which the subtraction becomes the
   addition of $-c$)
  and sets the rounding error for both operations to zero.
This particular improvement in rounding error handling
  is responsible for a dramatic reduction
  in false positives and false negatives
  inside standard library implementations of
  trigonometric functions like $\sin$ and $\cos$.

\subsection{Accurate Residue Functions}

The residue function for an operation $f(\hat{x}, \hat{y})$
  combines the rounding error $\mu_z$
  with the input residues $e_x$ and $e_y$
  to compute the residue $e_z$ of $z$.
We observed that the residue functions used by \eftsan
  contained several simplifications and inaccuracies.
Correcting these issues dramatically reduces
  false positives and false negatives
  on large-scale scientific software.

First, the \eftsan implementation included
  two typos in the residue functions.
The first bug affects the residue computation for subtraction,
  which should be $\mu_z + e_x - e_y$,
  but instead computes $\mu_z + e_x + e_y$.
The second, similar bug affects division,
  where the correct formula is
  $(e_x - \mu_z - \hat{z} \cdot e_y) / (\hat{y} + e_y)$,
  but where \eftsan adds instead of subtracting the $\mu_z$ term.
We reported both bugs to the \eftsan authors,
  who confirmed the issues;
  the \eftsan paper lists the correct (not buggy) formulas.
Fixing these bugs, especially the subtraction bug,
  leads to dramatic reductions
  in false positives and false negatives.
Since the paper in fact presents the correct formulas,
  all comparisons against \eftsan in this paper
  use a version with both typos corrected.

Second, \eftsan discards the ``higher-order error'' term
  in its residue function for multiplication.
That is, the residue $e_z$ in $\hat{z} = \hat{x} \cdot \hat{y}$
  should be
  $\mu_z + \hat{y} e_x + \hat{x} e_y + e_x e_y$,
  but \eftsan discards the $e_x e_y$ term.
Restoring it is especially important
  when $\hat{x}$ or $\hat{y}$ suffers from cancellation;
  in these cases $\hat{x} = 0$ but $e_x$ is nonzero.
If this affects both $\hat{x}$ and $\hat{y}$,
  the $e_x e_y$ term becomes the only nonzero term
  in the residue function,
  and dropping it would cause the debugger
  to lose track of the cancellations entirely.
Restoring the $e_x e_y$ term therefore also leads
  to a dramatic reduction in false positives and false negatives.

Third, for absolute value,
  \eftsan's residue function suffers from floating-point error.
The absolute value operation is exact,
  so it introduces no rounding error,
  but it must still have a residue function
  to propagate the effects of input error.
\eftsan computes the residue as
  $\lvert \hat{x} + e_x \rvert - \lvert \hat{x} \rvert$.
However, this is inaccurate when $e_x$ is small relative to $\hat{x}$:
  the $\hat{x} + e_x$ computation may round away the small residue $e_x$,
  causing the final residue to be incorrectly computed as $0$.
Instead,
  \name compares the signs of $\hat{x}$ and $\hat{x} + e_x$.
If both have the same sign,
  \name computes the residue as
  $\mathsf{copysign}(1, \hat{x}) \cdot e_x$.
When their signs differ,
  $e_x$ and $\hat{x}$ must have similar magnitude,
  and the \eftsan formula is used.
The effect of this change is smaller than that of the previous ones,
  but it still affects several benchmarks.

Finally, for square-root operations,
  \eftsan uses the residue function $(e_x + \mu_z) / (2\cdot \hat{z})$.
However, the $2\cdot \hat{z}$ term is actually an approximation
  of $\sqrt{\hat{x}} + \sqrt{\hat{x} + e_x}$,
  which \name uses instead;
  the two terms differ when $e_x$ is large.
We did not observe any changes in false positives or false negatives
  when switching to the more accurate formula in \name,
  but experience with the previous residue function corrections
  suggests that using the accurate residue functions is worthwhile.

\subsection{Overflow and Underflow}

The error-free transformations used by \name
  and previous machine-float residue debuggers
  do not apply when inputs are
  outside the normal numerical range.
\name therefore does not attempt to handle overflow and underflow,
  which in any case correspond to range errors
  rather than rounding errors.
To prevent overflow or underflow
  from causing false positives or false negatives
  for rounding errors,
  we guard rounding error estimation and residue function computation
  so that out-of-range inputs produce out-of-range outputs,
  and no warnings are generated for such values.

In our evaluation, we compare \name
  against an MPFR-based debugger.
MPFR has an expanded exponent range
  and can therefore represent values that overflow in machine precision.
To ensure a fair comparison,
  we modify the MPFR-based debugger
  to treat out-of-range values consistently with \name
  and to suppress warnings on overflow.

\subsection{Results on Scientific Benchmark Suites}

To demonstrate the accuracy impact
  of \name's rounding error estimation
  and residue function implementations,
  we evaluate it on three standard scientific software suites:
  the NAS Parallel Benchmarks (8 benchmarks)~\cite{npb},
  the Rodinia suite (18 benchmarks)~\cite{rodinia},
  and the Polybench suite (30 benchmarks)~\cite{polybench},
  discarding programs that do not use floating-point arithmetic
  (5 in Rodinia and 2 in Polybench).
We use a simple threshold where any residue
  greater than or equal to $2^{45}$ ULPs
  counts as a numerical warning%
\footnote{
  We chose this value to match \eftsan.
}
  and record the exact operations at which warnings are raised.
To count false positives and false negatives,
  we generalized \name to offer pluggable residue backends
  and implemented an MPFR backend
  with a benchmark-specific working precision
  between 128 and 4096 bits,
  to act as a ground truth.
We discard programs that run out of memory
  with MPFR even at 128 bits of precision (5 in total),
  leaving \numTotalBenchmarks~benchmarks.

\begin{figure}
  \includegraphics[width=\linewidth]{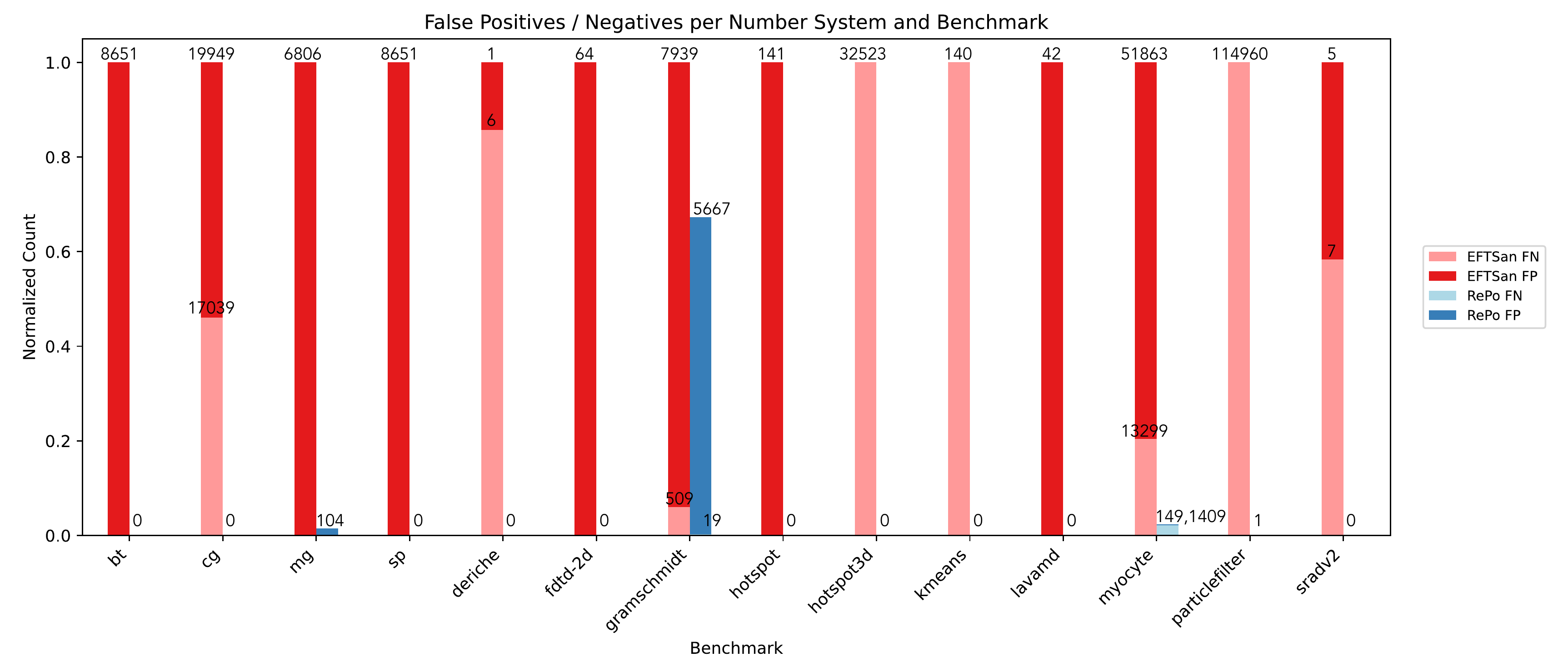}
  \caption{
  Number of false reports (false positives and false negatives)
    for the baseline residue algorithm
    (an \eftsan reimplementation with the subtraction and division bugs fixed)
    and our approach (\name).
  Only benchmarks with false positives or false negatives are shown;
    all bars are normalized so that the total bar height of \eftsan is 1.0.
    \name produces dramatically fewer false reports than \eftsan;
    only four benchmarks exhibit any false reports,
    and on three of them \name has orders of magnitude fewer.
  On the remaining benchmark, \texttt{gramschmidt},
    the false reports arise from residues clustering near
    the $2^{45}$ ULP threshold rather than from significant inaccuracies
    in the computed residues.
  }
  \label{fig:eval-sci}
\end{figure}

Overall, \name has false positives or negatives---false reports---%
  on just 4 of the \numTotalBenchmarks~benchmarks.
To compare to the current state of the art,
  we re-implemented \eftsan,
  with the subtraction and division bugs noted above fixed,%
\footnote{
  We also tested the original \eftsan
    subtraction and division residue functions but,
    as expected, they caused dramatically more false reports.
}
  as yet another pluggable backend.
The \eftsan reimplementation
  reports false positives or false negatives
  on \numSciInterestingBenchmarks~benchmarks,
  far more and a strict superset of those reported by \name;
  \Cref{fig:eval-sci} plots the results in detail.
Even when both debuggers report false positives or false negatives on a benchmark,
  \name reports far fewer.
On \texttt{particlefilter}, \name has just one false report,
  compared to over a hundred thousand for \eftsan,
  and on \texttt{mg} and \texttt{myocyte} it likewise produces
  orders of magnitude fewer.
Across all benchmarks,
  \name produces about $38\times$ fewer false reports than \eftsan,
  and on no individual benchmark does it produce more.
In fact, only on the \texttt{gramschmidt} benchmark
  do \name and \eftsan have
  even remotely similar false report counts:
  8\thinspace448 for \eftsan and 5\thinspace686 for \name.
We manually examined these false reports
  and found that many have a ground truth residue values
  close to the $2^{45}$ ULP threshold,
  suggesting that most of these remaining false reports
  come more from threshold effects
  rather than any remaining inaccuracy in the computed residues.
In short,
  \name achieves near-perfect results
  on standard scientific benchmark suites
  by carefully designing rounding error estimation
  and residue function implementations
  for maximum accuracy.
\section{Absorption and Residue Override}
\label{sec:override}

These results might suggest that
  accurately computing rounding errors and residue functions
  is sufficient for nearly perfect floating-point debugging.
While this may hold for
  already relatively accurate standard scientific benchmarks,
  it is far less true for challenging numerical cases
  due to an issue we name \emph{absorption}.
In fact, without proper treatment,
  not only can the \name debugger be inaccurate on such inputs
  (see \Cref{sec:eval}),
  but \emph{no} floating-point debugger
  can compute these residues accurately in a single execution
  without using prohibitively high precision.
  
\subsection{Absorption}

Absorption is the phenomenon
  that a numerical program may contain pairs of residues
  for which no single execution can measure both residues accurately.
This occurs when a residue $e_i$
  is influenced by two different rounding errors $\mu_{i^*}$ and $\mu_\ell$,
  and these rounding errors contribute to $e_i$
  with dramatically different magnitudes;
  for example, $\mu_{i^*}$'s contribution to $e_i$
  may be many orders of magnitude larger than $\mu_\ell$'s.
Since $e_i$ has limited precision,
  it cannot accurately represent both contributions.
A typical floating-point debugger therefore represents only the larger one,
  $\mu_{i^*}$,
  while rounding away the smaller one, $\mu_\ell$.
As a result, $e_i$ itself may be computed accurately.

However, some later residue $e_k$
  may combine $e_i$ with another residue $e_j$
  in such a way that the contribution of $\mu_{i^*}$ cancels,
  either with its own or with another rounding error
  $\mu_{j^*}$'s contribution to $e_j$.
In this case,
  $\mu_{i^*}$ and $\mu_{j^*}$ no longer contribute to $e_k$.
The smaller rounding error $\mu_\ell$
  may still contribute to $e_k$,
  but because it was not represented in $e_i$,
  that contribution is lost,
  leaving $e_k$ computed inaccurately.

This observation leads to an \emph{impossibility} result:
  \emph{no} floating-point debugger
  that stores residues using machine floating-point values
  can accurately compute all residues in all situations,
  especially in challenging numerical code
  involving multiple cancellations
  or the internals of library functions.

At its core, absorption is simply a consequence
  of limited precision in representing residues,
  and may initially appear unsurprising.
What \emph{is} surprising, however,
  is that it is often possible to measure $e_k$ accurately,
  though at the cost of measuring $e_i$ inaccurately.
Imagine that, instead of storing $\mu_{i^*}$'s large contribution,
  $e_i$ stored only the smaller contribution $\mu_\ell$.
This would make $e_i$ highly inaccurate,
  but it would preserve $\mu_\ell$'s contribution
  for the later computation of $e_k$.
In many challenging numerical programs,
  this would allow $e_k$ to be computed accurately.

In short, \emph{both} $e_i$ and $e_k$
  can be computed accurately in this scenario,
  but not within the same execution.
This observation suggests a way around the impossibility result:
  perform multiple executions,
  each measuring different subsets of residues.
We call this solution \RO.

\subsection{Residue Override}

\begin{figure}
  \centering
  \includegraphics[scale=0.25]{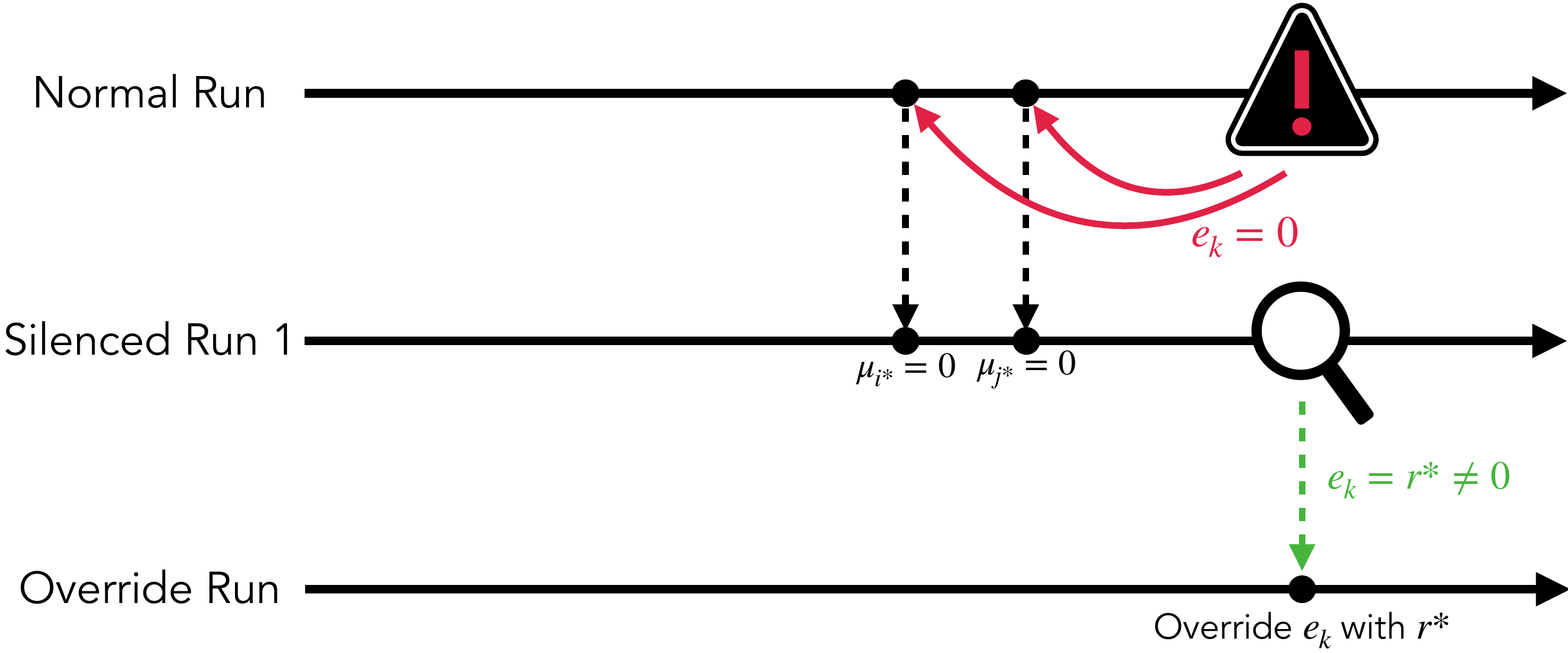}
  \caption{The \RO framework
    estimates more precise residue values
    by executing the target program three times in total.
  The first execution \emph{detects}
    cancellation ($e_k = 0$)
    and identifies the leading terms that cancel
    ($i^*$ and $j^*$).
  The second execution silences operations $i^*$ and $j^*$,
    ignoring their rounding errors
    when computing residues ($\mu_{i^*} = \mu_{j^*} = 0$).
  This results in a more accurate, non-zero value $r^* \ne 0$
    for the residue $e_k$.
  The third execution then re-enables $i^*$ and $j^*$,
    but overrides the residue $e_k$ with $r^*$,
    replacing the cancellation with a more accurate residue value.
  }
  \label{fig:ro-simple}
\end{figure}

The basic idea of \RO is illustrated in \Cref{fig:ro-simple},
  which uses the same assumption introduced earlier:
  residue $e_k$ suffers from cancellation when it combines
  residues $e_i$ and $e_j$ such that their largest contributors cancel.
The method consists of three executions of the target program.
The first two executions measure different sets of residues,
  and the last execution combines them.

In the figure, the first execution, labeled ``Normal Run'',
  measures residues $e_i$ and $e_j$ accurately,
  but as a result cannot measure $e_k$ accurately.
The second execution, labeled `Silenced Run 1'' in the figure,
  \emph{silences} rounding errors $\mu_{i^*}$ and $\mu_{j^*}$,
  excluding their contributions when computing $e_i$ and $e_j$.
This means that the silenced run
  measures $e_i$ and $e_j$ inaccurately,
  but may compute an accurate value for $e_k$,
  which the figure labels $r^*$.
We say that this second run also \emph{probes} $e_k$,
  meaning that it records the value that $e_k$ takes in this run.
Note that silencing $\mu_{i^*}$ and $\mu_{j^*}$
  affects only the debugger-computed residues,
  not the actual program values,
  and therefore cannot change control flow or other program behavior.
The third execution, labeled ``Override Run'' in the figure,
  no longer silences $\mu_{i^*}$ and $\mu_{j^*}$,
  and thus again measures $e_i$ and $e_j$ correctly.
However, instead of computing $e_k$ normally---%
  and thus inaccurately!---%
  this final run \emph{overrides} its value with $r^*$,
  the value computed during the silenced run.
In other words, the final override run
  obtains accurate values for $e_i$, $e_j$, \emph{and} $e_k$.

Making this basic idea work requires a critical step:
  \emph{detecting} when absorption occurs,
  in which case $e_k$ must be estimated separately from $e_i$ and $e_j$
  and can be measured accurately only by silencing
  the dominant rounding errors $\mu_{i^*}$ and $\mu_{j^*}$.
\name achieves this through a simple mechanism.
Every floating-point operation executed during the program
  is assigned an ``operation ID'',
  which is just an auto-incrementing 64-bit number.
The program is assumed to be deterministic,
  so these operation IDs serve as stable identifiers
  across multiple executions.

The debugger stores,
  for every floating-point value $i$ in the program,
  not only its residue $e_i$
  but also the operation ID $i^*$
  of the rounding error $\mu_{i^*}$
  that makes the largest contribution to $e_i$.
When an operation $\hat{z} = \hat{f}(\hat{x}, \hat{y})$ computes a residue $e_z$,
  \name monitors the numerical error of the residue computation.
If large numerical error is detected,
  absorption is assumed to prevent
  the output residue $e_z$ and
  the input residues $e_x$ and $e_y$
  from simultaneously being computed accurately.
The largest contributors $\mu_{i^*}$ and $\mu_{j^*}$
  to $e_x$ and $e_y$
  are then silenced on the next run.
Silencing the largest contributors to $e_x$ and $e_y$
  makes those residues less accurate,
  but also allows smaller contributors to $e_x$ and $e_y$
  to be represented,
  which may enable $e_z$ to be computed more accurately.%
\footnote{
Of course, more complicated situations can arise;
  these are discussed in \Cref{sec:impl}.
}

Formally, every residue function in \name
  for an operation $\hat{z} = \hat{f}(\hat{x}, \hat{y})$
  is structured as $e_z = A \mu_z + B e_x + C e_y$,
  where $A$, $B$, and $C$ can depend on
  $\hat{x}$, $\hat{y}$, and their residues $e_x$ and $e_y$.
Since $A$, $B$, and $C$ can depend on $e_x$ and $e_y$,
  this is not a purely ``linear'' or ``first-order'' residue function.
For example, for a multiplication operation $\hat{z} = \hat{x} \cdot \hat{y}$,
  we define
\[
e_z = 1 \cdot \mu_z
+ \left(\hat{y} + \frac12 e_y\right) e_x
+ \left(\hat{x} + \frac12 e_x\right) e_y;
\]
  this expands to the expected
  $\mu_z + \hat{y} e_x + \hat{x} e_y + e_x e_y$,
  which includes a higher-order error term.

Separating the $A \mu_z$, $B e_x$, and $C e_y$ terms
  makes it easy to define the ``largest contributor'' to $e_z$.
We compare the magnitudes of these three terms.
If $|A \mu_z|$ is the largest,
  then $\mu_z$ is the largest contributor to $e_z$.
If $|B e_x|$ or $|C e_y|$ is largest,
  then the largest contributor to $e_z$
  is the largest contributor to $e_x$ or $e_y$, respectively.
In the case of ties, the order is $e_z$, $e_x$, and then $e_y$.
Through this mechanism,
  every residue has an assigned largest contributor.%
\footnote{
For exactly computed values,
  which have no contributing rounding errors,
  a dummy largest contributor ($-1$) is assigned.
}

If high numerical error is detected while computing $e_z$,
  the largest contributors to $e_x$ and $e_y$
  are then silenced and $e_z$ is probed on the next run.
A final override run then combines these measurements
  to provide accurate residues for floating-point debugging.
This \emph{detect}-\emph{silence}-\emph{probe}-\emph{override} sequence
  is effective at reducing false positive and false negative warnings
  on a number of challenging numerical benchmarks,
  as discussed in \Cref{sec:eval}.
We now turn to the question
  of how \name implements this basic algorithm.   
\section{Implementing Residue Override}
\label{sec:impl}

\Cref{sec:override} illustrated the basic idea of \RO;
  we now describe how to implement it in practice.

\subsection{Tracking the Largest Error Contributor}

\Cref{sec:override} describes how to identify the largest contributor
  to a residue by expressing
\[
e_z = A \mu_z + B e_x + C e_y
\]
and comparing the magnitudes of the three terms.
To implement this mechanism,
  \name stores the identity of the largest contributor
  in a field \texttt{maxErrOp} within each residue.

Let \texttt{absIntroErr} denote $|A \mu_z|$,
  \texttt{absAmpErr1} denote $|B e_x|$,
  and \texttt{absAmpErr2} denote $|C e_y|$.
The \texttt{maxErrOp} field is assigned as follows:

\begin{lstlisting}[language=C,style=paperlisting]
if (absIntroErr >= max(absAmpErr1, absAmpErr2))
    e_z->maxErrOp = curOp;
else if (absAmpErr1 >= absAmpErr2)
    e_z->maxErrOp = e_x->maxErrOp;
else
    e_z->maxErrOp = e_y->maxErrOp;
\end{lstlisting}

Thus, if the local rounding error dominates,
  the current operation is recorded;
otherwise, the dominant contributor is inherited
  from the input with the larger propagated error.
Each residue therefore carries the identity of the operation
  that contributes most to its error.

\subsection{Detection of Absorption}
In the \RO framework, the detection phase determines when
  residue override should be triggered.
The silence, probe, and override phases are largely self-contained
  and straightforward to implement,
  but the detection step requires more care.

To identify cases where the residue computation suffers
  from significant numerical error,
  we must determine whether a near-zero residue
  is the result of cancellation that hides a larger error.
Simply checking whether the residue value is small is insufficient:
  we do not want to trigger \RO when a residue is zero
  because the computation is genuinely well behaved
  (for example, when a cancellation is mathematically expected
  and no absorption occurs).
Formally, let the residue $e_z$ be computed by the residue function
  $e_f$.
\name aims to trigger detection only when both
\[
  e_z \approx 0
  \qquad\text{and}\qquad
  \textsc{nontrivial}(e_f)
\]
hold.
To evaluate these conditions, \name employs two heuristics.

The first condition, $e_z \approx 0$, captures potential cancellations.
An exactly zero residue is trivial to detect,
  but it is also useful to detect cases
  where the residue is very small but nonzero due to cancellation.
To this end, \name adds an \texttt{isZero} flag to each residue.
This flag is set when either the residue is exactly zero or
  the condition number of the residue function,
\[
  \frac{|A\mu_z| + |B e_x| + |C e_y|}{|e_z|},
\]
exceeds a predefined large threshold,
  where $e_z = A\mu_z + B e_x + C e_y$.
Intuitively, if the contributing terms nearly cancel
  so that $|e_z| \ll |A\mu_z| + |B e_x| + |C e_y|$,
  the ratio becomes large and the residue is flagged.
This idea is similar to ATOMU's notion of atomic conditions~\cite{atomu},
  except that it applies to the debugger's shadow computations
  rather than to the program's original operations.

The second condition, $\textsc{nontrivial}(e_f)$,
  distinguishes harmful cancellations from benign ones.
Residue computations may cancel even when no rounding error
  has actually rounded away any error terms.
Such cancellations are numerically correct,
  and residue override would be unnecessary.
To avoid triggering on these benign cases,
  \RO introduces an \texttt{isAbsorbed} flag for each residue.
The \texttt{isAbsorbed} flag is determined by comparing
  the largest error contribution
  (which determines \texttt{maxErrOp})
  to the final residue value.
If the largest contribution accounts for an overwhelming fraction---%
  all but a few ULPs---%
  of the final residue, \texttt{isAbsorbed} is set.
When a cancellation occurs between two residues
  whose \texttt{isAbsorbed} flags are both clear,
  \RO treats the cancellation as benign
  and does not trigger override.
\RO triggers only when a residue has both
  its \texttt{isZero} and \texttt{isAbsorbed} flags set,
  after which \name identifies the leading contributors
  $i^*$ and $j^*$ to the two input residues
  and initiates the silence-probe-override procedure.

Both the cancellation condition and the nontriviality condition
  involve thresholds that must be set heuristically.
Empirically, we find that adjusting these thresholds trades off
  between the number of re-executions that \name performs
  and the number of residues that \name can correct using \RO.
We leave more principled approaches for identifying
  absorption-affected residues to future work;
  nevertheless, this simple heuristic approach
  already yields good results in \Cref{sec:eval}.

\section{Complex Absorption Scenarios}
\label{sec:pro}

The implementation described in \Cref{sec:impl}
  resolves the simplest case of absorption:
  a single inaccurately computed residue
  that can be corrected by silencing one pair of operations.
However, real-world numerical applications
  often exhibit more complex absorption patterns.
To make \RO practical in such settings,
  \name must address two additional challenges:
  absorptions that require silencing multiple pairs of operations,
  and programs that exhibit multiple independent absorptions.
The solution builds on the same core \RO steps---%
  detect, silence, probe, and override---%
  but requires interleaving these stages.

\subsection{Handling Multi-Term Absorption}

\begin{figure}
  \centering
  \includegraphics[width=\linewidth]{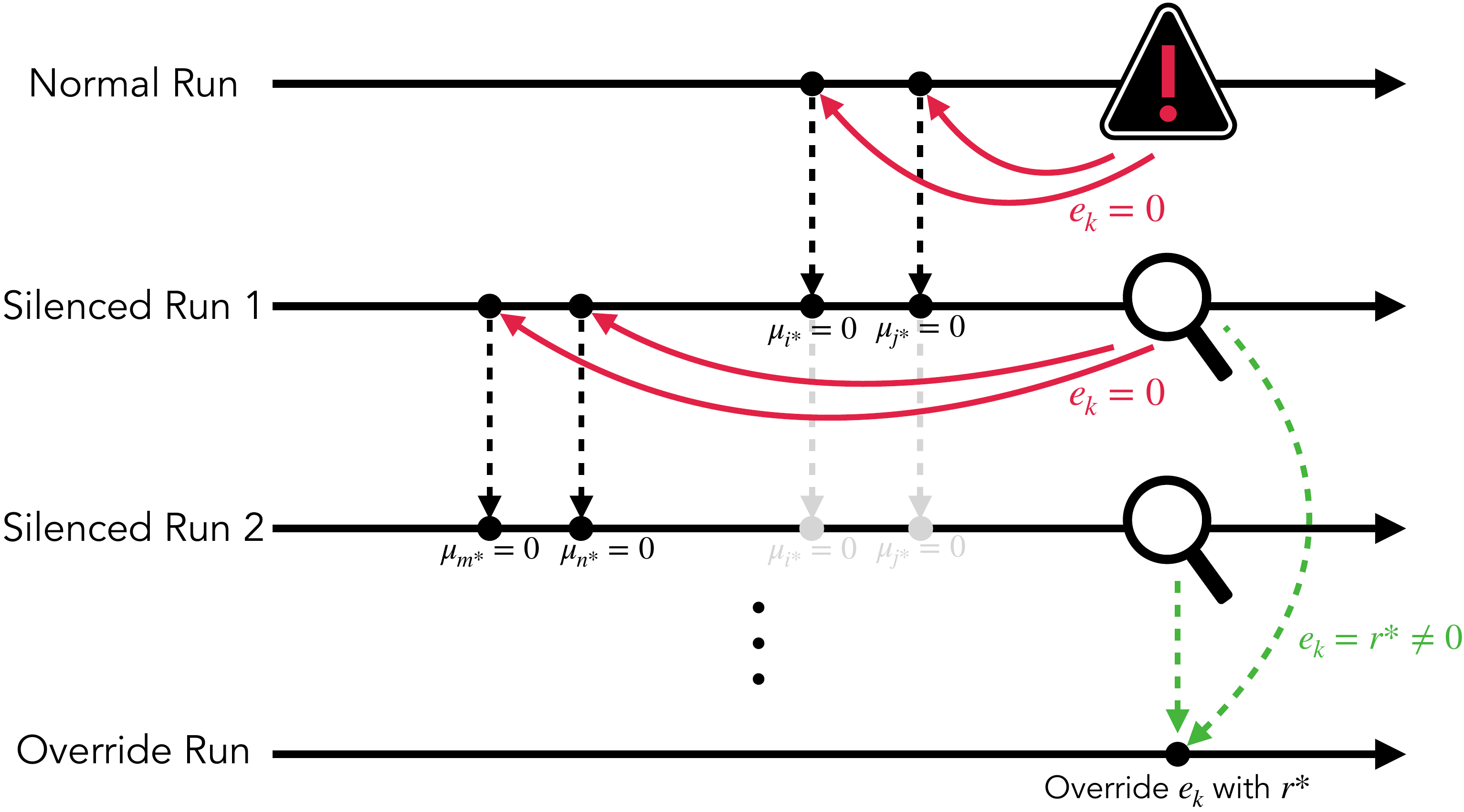}
  \caption{
    Repeated silencing in \name.
    Probing the first silenced run
      still flags $e_k$ as an absorption
      and thus identifies two additional largest contributors, $m^*$ and $n^*$.
    \name then performs a second silenced run
      silencing all four identified largest contributors.
    Eventually, a nonzero value $r^*$
      is obtained for the residue $e_k$,
      allowing \name to perform the final override run
      with the corrected value $r^*$.
  }
  \label{fig:ro-complex}
\end{figure}

\Cref{fig:ro-complex} illustrates an absorption scenario in which
  absorption is detected for residue $e_k$,
  the two largest contributors $i^*$ and $j^*$ are silenced,
  but, when probed, $e_k$ is still detected to suffer from absorption.
This occurs because absorption can affect
  arbitrarily large sets of mutually immeasurable residues.
Such patterns are particularly common
  in complex routines, such as implementations of $\sin$.

To handle these cases, \name continues to track
  the largest contributor to each residue
  \emph{even during the silenced run},
  and correspondingly updates the \texttt{maxErrOp} field.
In the silenced run, some rounding errors are set to zero,
  and are therefore no longer the largest contributors
  to any residue,
  even if they were during the initial run.
Thus, if absorption is detected at $e_k$ during the silenced run,
  two \emph{additional} operations $m^*$ and $n^*$
  are identified as contributing to the absorption.
\name then performs \emph{another} silenced run
  in which $i^*$, $j^*$, $m^*$, and $n^*$ are all silenced,
  and, if absorption is still detected,
  further runs silencing additional operations can be performed.
This process stops when $e_k$ is no longer detected as suffering from absorption,
  meaning the \texttt{isZero} or \texttt{isAbsorbed} flag of $e_k$
  is no longer set.

\subsection{Handling Multiple Independent Absorptions}

\begin{figure}
  \centering
  \includegraphics[width=\linewidth]{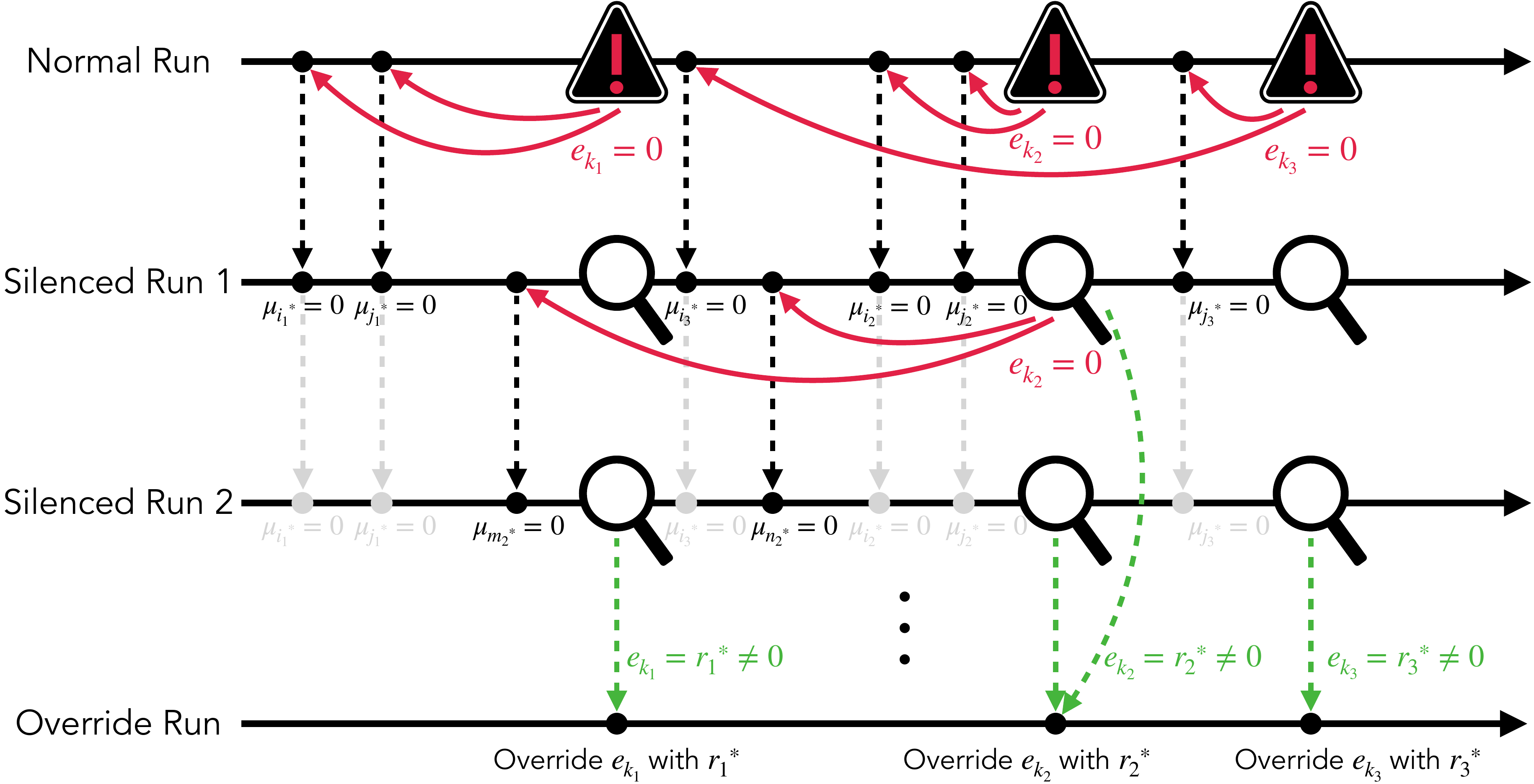}
  \caption{
    Handling multiple absorptions simultaneously.
    During the initial run,
      three residues ($k_1$, $k_2$, and $k_3$) exhibit absorption.
    In subsequent silenced runs,
      \RO successively silences the dominant contributors ($i^*$, $j^*$, etc.)
      and recomputes the affected residues
      until each residue obtains a nonzero value (${r_1}^*$, ${r_2}^*$, ${r_3}^*$).
    Finally, in the override run,
      \RO replaces the corresponding residues $e_{k_1}$, $e_{k_2}$, and $e_{k_3}$
      with the corrected values ${r_1}^*$, ${r_2}^*$, and ${r_3}^*$.
  }
  \label{fig:pro}
\end{figure}

Multi-term absorptions still affect only a \emph{single}
  residue in a program.
However, in real-world numerical software,
  there are often \emph{multiple} residues
  affected by absorption.
These residues may be independent of one another
  (discussed in this subsection)
  or, worse, may interact (discussed in the next one).
When the residues are independent,
  \name attempts to resolve all absorptions
  within the same re-execution.

At a high level, \name does so by performing
  multiple silences and probes simultaneously,
  as shown in \Cref{fig:pro}.
In the first execution of \name,
  three residues are affected by absorption:
  $k_1$,~$k_2$, and~$k_3$.
For each one, \name identifies its largest contributors:
  $i_1^*$,~$j_1^*$ for $k_1$,
  $i_2^*$,~$j_2^*$ for $k_2$,
  and~$i_3^*$,~$j_3^*$ for $k_3$.
In the next run,
  all six largest contributors are silenced,
  and more accurate values of $k_1$, $k_2$, and $k_3$ are probed.
It is possible that some of the probed residues
  are still affected by absorption, such as $k_2$ in the figure;
  in this case, \name silences the next pair of largest contributors,
  as in the case of multi-term absorptions.
Finally, once all probed values are accurate,
  \name proceeds to the override stage.

Practically, to handle multiple absorptions,
  \name maintains four data structures across runs:
\begin{itemize}
  \item \texttt{silentOps}, the set of operations whose rounding errors
    should be silenced;
  \item \texttt{probeOps}, the set of operations whose residues
    should be probed;
  \item \texttt{tempResOverride}, a temporary mapping
    from probed operations $i$ to their newly computed residues $r_i^*$;
  \item \texttt{resOverride}, a permanent mapping
    from operations to their accurately computed residues.
\end{itemize}
All four data structures are stored on disk,
  loaded by \name at the beginning of each re-execution,
  and used and updated according to this logic:

\begin{lstlisting}[language=C,style=paperlisting]
void compute_residue(shadow_value_t *e_z) {
  if (resOverride.contains(curOp)) { *e_z = resOverride[curOp]; }
  if (silentOps.contains(curOp)) { introErr = 0; }
  else {
    // Normal error-free transformation routine
  }
  // Compute e_z
  if (probeOps.contains(curOp)) { tempResOverride[curOp] = e_z->value; }
}
\end{lstlisting}

Between re-executions,
  operations involved in absorptions are added to \texttt{silentOps} and \texttt{probeOps};
  entries in \texttt{probeOps} are moved to \texttt{tempResOverride};
  and entries in \texttt{tempResOverride} are moved to \texttt{resOverride}
  if no additional absorptions are detected.
This scheme can efficiently handle multiple absorptions
  in as few as three executions.

\subsection{Second-Largest Contributor for Parallel Resolution}

Finally, when multiple residues suffer from absorption,
  it is also possible for those residues to interact with one another.
The first, simpler case of interaction
  is when accurately computing one residue
  causes later residues to suffer from absorption.
This is common, because
  residues that suffer from absorption are typically very small,
  and computing those residues more accurately
  often makes them larger,
  meaning that later residues may acquire new, dominant contributors.
\name detects this case when,
  during a silence or override run,
  a residue that did not \emph{previously} suffer from absorption
  begins to suffer from it.
In this case,
  \name adds the new residue's largest contributors
  to the silenced set,
  the new residue to the probed set,
  and performs another re-execution.

The more challenging case of interaction
  is when two different residues both suffer from absorption,
  but silencing one residue's largest contributor
  affects the probed value of the other residue.
For example, an important, but not largest,
  contributor to one absorption may also be
  the largest contributor to another absorption.
Silencing this contributor (to probe the second absorption)
  would remove an important contribution for the first absorption,
  causing the probe of the first absorption to measure an incorrect value.

In this case, the two residues cannot be
accurately measured in the same re-execution.
\name therefore detects this issue
and measures them in separate re-executions.
To detect this case,
\name tracks the \emph{second-largest} contributor to each residue,
denoted \texttt{sndErrOp}.
It is computed
together with the largest contributor \texttt{maxErrOp}, as follows:

\begin{lstlisting}[language=C,style=paperlisting]
if (absIntroErr >= max(absAmpErr1, absAmpErr2)) {
    e_z->sndErrOp = (absAmpErr1 > absAmpErr2) ? e_x->maxErrOp : e_y->maxErrOp;
}
else if (absAmpErr1 >= absAmpErr2) {
    e_z->sndErrOp = (absIntroErr > absAmpErr2) ? curOp : e_y->maxErrOp;
}
else {
    e_z->sndErrOp = (absIntroErr > absAmpErr1) ? curOp : e_x->maxErrOp;
}
\end{lstlisting}

The key distinction between
  the largest and second-largest contributors
  is that the largest contributor determines
  \emph{which} operation to silence,
  while the second-largest contributor determines
  \emph{whether} two absorptions can be resolved simultaneously.
When multiple absorptions are detected,
  \name iterates through them,
  adding the largest contributors of each absorption
  to the silenced set
  \emph{only} if their second-largest contributors
  are not already present in the silenced set.
After a successful silencing-and-probing run,
  some absorptions are resolved,
  their largest contributors are removed from the silenced set,
  and later interacting residues are then
  silenced and probed.

\subsection{Summary}

The final data structure of a residue in \name is as follows:

\begin{lstlisting}[language=C,style=paperlisting,xleftmargin=0pt]
struct Residue {
  double value; // computed residue
  OpId maxErrOp; // largest contributor operation ID
  OpId sndErrOp; // second-largest contributor operation ID
  bool isAbsorbed; // absorption flag
  bool isZero; // zero flag
};
\end{lstlisting}

In other words, each residue stores
  its computed value,
  its largest and second-largest contributor operations,
  and its two flags.
These residues are then organized using the
  \texttt{silentOps}, \texttt{probeOps}, and \texttt{resOverride} sets,
  which are updated as described in \Cref{alg:pro}.
This algorithm enables \name to resolve
  complex, real-world absorption scenarios
  with as few re-executions as possible.

\begin{algorithm}
\caption{Parallel Residue Override}\label{alg:pro}
\sf \footnotesize
\begin{algorithmic}

\Procedure{resolve}{p, silentOps, probeOps, resOverride}
\State \textbf{repeat}
  \State stillCancel $\gets$ \textbf{false}
  \State maxErrOps, sndErrOps $\gets \emptyset$
  \State tempResOverride, absorptions $\gets$ \textsc{execute}(p, silentOps, probeOps, resOverride)
  \For{($i_x^*$, $j_x^*$, $i_y^*$, $j_y^*$, $k$) $\in$ absorptions}
  \If{$(k \in \texttt{probeOps}) \land ({i_x^*, i_y^*} \cap \texttt{sndErrOps} = \emptyset) \land ({j_x^*, j_y^*} \cap \texttt{maxErrOps} = \emptyset)$}
  \State stillCancel $\gets$ \textbf{true}
  \State silentOps.\textsc{add}($i_x^*, i_y^*$)
  \State maxErrOps.\textsc{add}($i_x^*, i_y^*$)
  \State sndErrOps.\textsc{add}($j_x^*, j_y^*$)
  \EndIf
  \EndFor
  \State \textbf{until} not stillCancel
\State \Return tempResOverride
\EndProcedure

\Procedure{RePo}{p} \Comment{driver program}
\State hasCancel $\gets$ \textbf{false}
\State silentOps, probeOps, resOverride $\gets \emptyset$
\State maxErrOps, sndErrOps $\gets \emptyset$

\While{\textbf{true}}
\State \_, absorptions $\gets$ \textsc{execute}(p, silentOps, probeOps, resOverride)
\State hasCancel $\gets$ \textbf{false}

\For{($i_x^*$, $j_x^*$, $i_y^*$, $j_y^*$, $k$) $\in$ absorptions}
  \If{$(\{i_x^*, i_y^*\} \cap \texttt{sndErrOps} = \emptyset) \land (\{j_x^*, j_y^*\} \cap \texttt{maxErrOps} = \emptyset)$}
    \State hasCancel $\gets$ \textbf{true}
    \State silentOps.\textsc{add}($i_x^*, i_y^*$)
    \State probeOps.\textsc{add}($k$)
    \State maxErrOps.\textsc{add}($i_x^*, i_y^*$)
    \State sndErrOps.\textsc{add}($j_x^*, j_y^*$)
  \EndIf
\EndFor

\If{not hasCancel}
  \State \textbf{break}
\EndIf

\State tempResOverride $\gets$ \textsc{resolve}(p, silentOps, probeOps, resOverride)
\State resOverride.\textsc{update}(tempResOverride)
\State \textsc{clear}(silentOps, probeOps, maxErrOps, sndErrOps)

\EndWhile
\State \Return
\EndProcedure

\end{algorithmic}
\end{algorithm}

\section{Evaluation}
\label{sec:eval}

Even without \RO, \name is substantially more accurate
  than existing state-of-the-art debuggers
  on standard scientific benchmark suites,
  and \RO rarely triggers on them.
To evaluate the impact of \RO,
  we therefore focus on more challenging numerical benchmarks.
We aim to answer three research questions:

\begin{enumerate}
\item[RQ1:] Does \RO improve \name's false report rate?
\item[RQ2:] How many re-executions does \RO require?
\item[RQ3:] How does \RO compare to high-precision arithmetic?
\end{enumerate}

We draw challenging numerical programs from two suites:
  \eftsan's \numMicrobench-program test suite~\cite{eftsan},
  which is drawn from prior floating-point debugging work,
  and the standard FPBench~\cite{fpbench} suite,
  which contains 130~numerical kernels
  collected from numerical analysis papers and textbooks.
The FPBench benchmarks are expressions rather than complete programs,
  so we compile them into C functions using the FPBench toolchain
  and generate simple driver programs to invoke these functions
  with random inputs (using a fixed seed for determinism).%
\footnote{
We exclude the \texttt{apron} suite,
  which the FPBench toolchain cannot compile to C.
Each driver generates 500 inputs
  spanning a range of exponents
  and executes the kernel on all 500 inputs in a loop.
}
Six benchmarks run out of memory
  when computing ground-truth MPFR residues,
  leaving a total of 169 benchmarks.
We focus on a subset of \numMFInteresting~benchmarks
  for which the first run of \name produces false reports.
These benchmarks represent the most challenging cases
  and are also those where \RO is expected to have an effect.%
\footnote{The bug-fixed re-implementation of \eftsan
  produces false reports on all \numMFInteresting~benchmarks
  as well as 18 additional benchmarks.}
Several benchmarks involve
  math library functions such as \texttt{sin}, \texttt{exp}, or \texttt{log},
  whose implementations are particularly challenging to analyze.
We cap \name's \RO algorithm at 20 re-executions
  and perform all experiments on a machine with
  Ubuntu~24.04.3, an Intel~i7-8700K at 3.70\,GHz, and 32\,GB of memory,
  using Clang~18.1.3 for all compilation.

\paragraph{RQ1}

\begin{figure}
  \includegraphics[width=\textwidth]{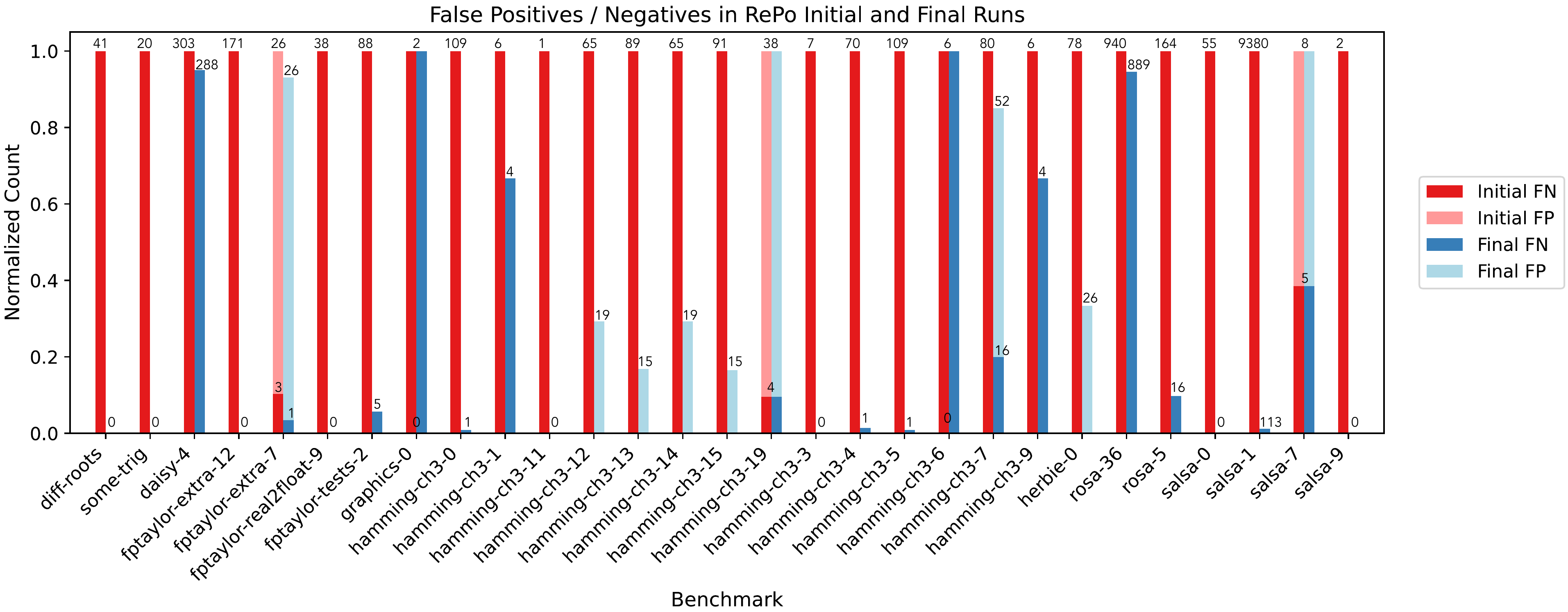}
  \caption{
  Number of false reports (false positives and false negatives)
    for the initial and final runs of \name.
  Only benchmarks for which \name has non-perfect initial results are shown;
    all bars are normalized so that the total height of the initial run is 1.0.
  \name produces fewer false reports on \numROTotalImproved~out of \numROInteresting~benchmarks.
  On 19 of these, the \RO mechanism reduces the number of false negatives
    by about $10\times$ compared to the initial run.
  }
  \label{fig:rq1}
\end{figure}

Among the \numMFInteresting~benchmarks,
  \RO triggers on \numROInteresting~of them.
\Cref{fig:rq1} shows their initial and final false reports.
On \numROTotalImproved~benchmarks (\pctROTotalImproved),
  \RO reduces the total number of false reports,
  often dramatically.

Among these \numROTotalImproved~benchmarks,
  \RO improves both false positives and false negatives
  on \numROStrictImproved~of them,
  reducing at least one without increasing the other.
Specifically,
  \numROPerfect~of them achieve perfect results (no false reports)
  in their final runs.
For example, the \texttt{diff-roots} benchmark,
  which is used as the running example in \Cref{sec:overview},
  evaluates $\sqrt{x + 1} - \sqrt{x}$ on 100 positive inputs with varying exponents.
\name's initial run produces 41 false negatives,
  but \RO detects and corrects all inaccurate residues,
  achieving perfect results in the final override run.
Moreover, some benchmarks would achieve perfect results
  with a higher cap on re-executions.
For example, \texttt{hamming-ch3-1} in the FPBench suite,
  which computes $\sin(x + \epsilon) - \sin(x)$,
  resolves two of six false negatives within the first 20 re-executions,
  but resolves the remaining four false negatives
  with just three additional re-executions.
This benchmark requires many re-executions
  because it must handle complex absorptions arising
  from multi-precision arithmetic inside the $\sin$ implementation.

For \numROMixed~out of \numROTotalImproved~benchmarks,
  \RO reduces the number of false negatives
  but introduces some false positives.
For example, on the \texttt{hamming-ch3-14} benchmark,
  \RO eliminates all 65 false negatives from the first run
  but introduces 19 false positives.
Even so, in these cases,
  \RO reduces both the total number of false reports
  and, in particular, the number of false negatives,
  which are the most dangerous type of error for a debugger.

On \numRONoHelp~benchmarks,
  residue override does not change the number of false reports.
However, for two of them, \texttt{graphics-0} and \texttt{hamming-ch3-6},
  this is due to the re-execution cap; increasing the cap
  allows both benchmarks to achieve perfect results
  but requires 50 re-executions.
We expect that future work can examine more efficient methods
  for reasoning about elementary function implementations
  that reduce the number of required re-executions.

In summary, \textbf{\name's \RO mechanism significantly reduces false reports}
  even on the most challenging numerical benchmarks.

\paragraph{RQ2}

\begin{figure}
  \centering
  \includegraphics[width=\linewidth]{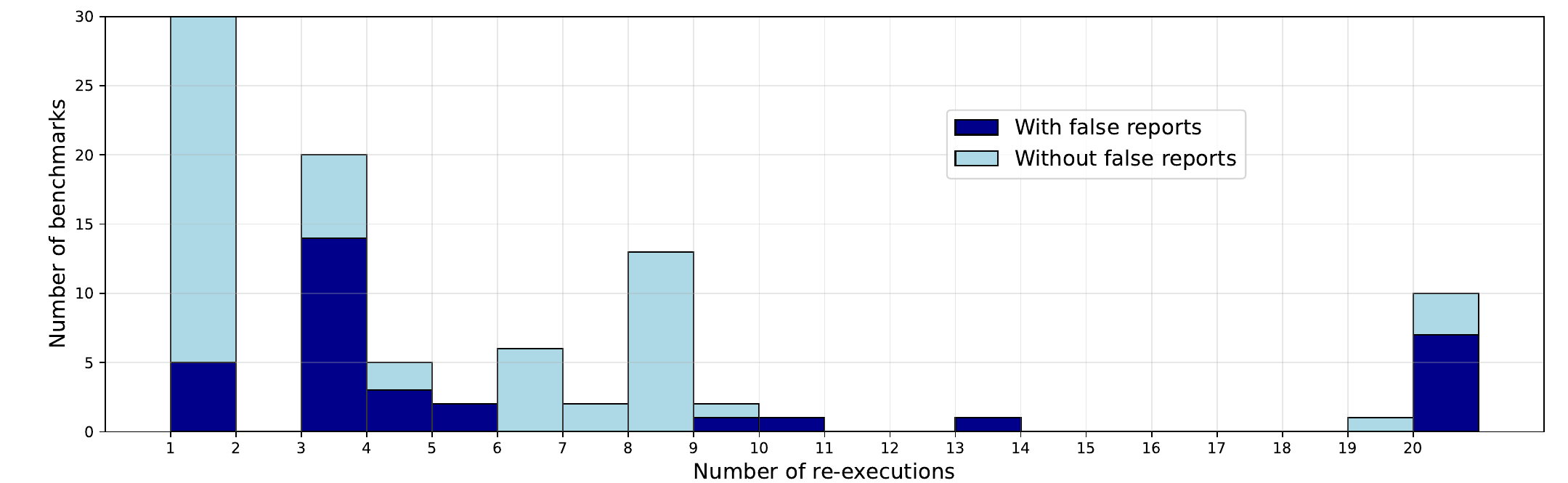}
  \caption{
  Number of re-executions for all \numMFBenchmarks~benchmarks
    (34 benchmarks with false reports in their initial runs, and
    135 benchmarks without false reports in their initial runs).
  We cap the plot at 30 benchmarks
    to emphasize benchmarks whose initial runs have false reports.
  For benchmarks without false reports (already accurate),
    the distribution of re-executions is highly right-skewed:
    101 out of 135 benchmarks require no re-executions.
  For benchmarks with false reports,
    \RO is triggered on most of them,
    and nearly half require only three executions,
    thanks to its ability to handle multiple absorptions simultaneously.
  }
  \label{fig:rq2}
\end{figure}

While \RO is effective, it requires multiple executions of the program,
  introducing additional debugging overhead.
We therefore measure how many re-executions are required
  for \name's \RO algorithm to terminate.
\Cref{fig:rq2} shows the results as a histogram.

Across all \numMFBenchmarks~benchmarks,
  \name requires \avgReexecutions~re-executions on average,
  with \numMFNonRO~benchmarks requiring no re-executions
  because no absorption is detected.
For the remaining 63 benchmarks,
  \name requires at least 3 re-executions:
  one for the initial detection run,
  one for the silenced run,
  and one for the final override run.
Twenty benchmarks require only these 3 re-executions,
  while the remaining benchmarks require more,
  often due to calls to library functions such as \texttt{sin},
  which introduce deeper chains of cancellation
  and therefore require additional re-executions.
Ten benchmarks reach the cap of \capReexecutions~re-executions.

Benchmarks \emph{without} false reports in their initial runs
  require very few re-executions in \Cref{fig:rq2}:
  on these benchmarks, \name requires 2.7 re-executions on average;
  in contrast, on the \numMFInteresting~benchmarks considered in RQ1,
  where the initial runs produce false reports,
  it requires 7.1 re-executions.
This highlights an advantage of \RO
  over higher-precision floating-point debuggers:
  for already accurate applications,
  \name incurs very low overhead.
We explore \name's overhead further in RQ3.

Some benchmarks trigger the residue override mechanism
  but produce the same final results as the initial run.
This occurs when the detected absorptions
  do not affect whether the computed residues
  cross the reporting threshold.
While these cases do not improve the reported results,
  more sophisticated floating-point debuggers
  may still benefit from the improved residue accuracy.

In summary,
  \textbf{\name requires an acceptable number of re-executions},
  even for challenging numerical benchmarks.

\paragraph{RQ3}

\begin{figure}
  \centering
  \includegraphics[width=\linewidth]{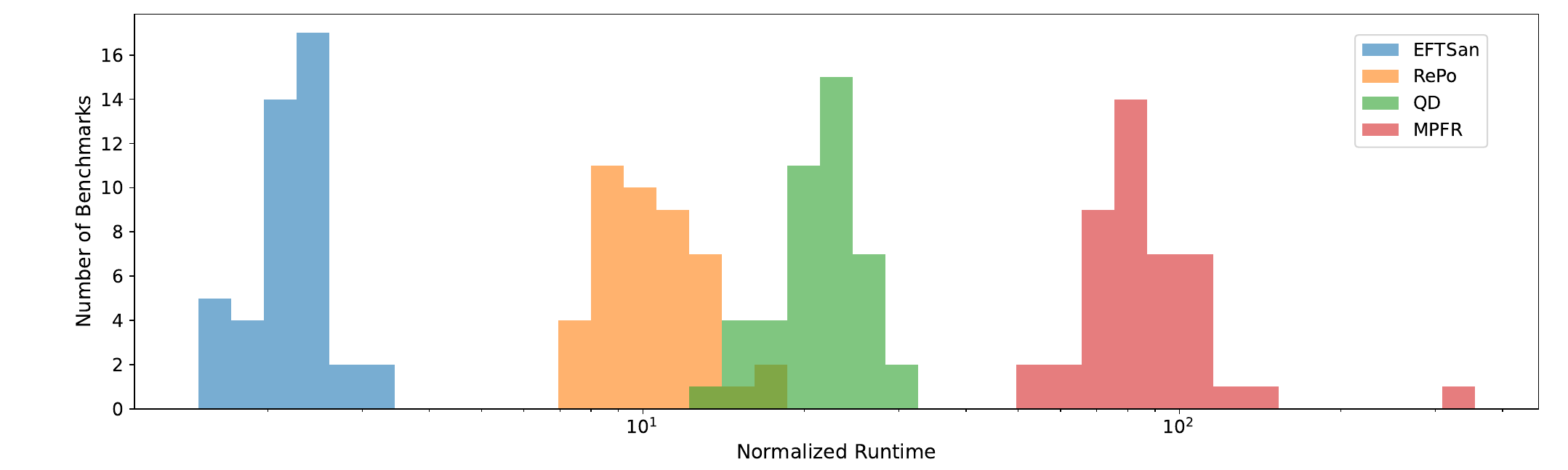}
  \caption{
  Distribution of runtime overhead for
    \eftsan, \name, QD, and MPFR,
    normalized to uninstrumented execution.
  The $x$-axis is logarithmic to capture
    the wide range of overheads.
  Overall, \name exhibits overhead
    somewhat higher than \eftsan,
    generally lower than QD,
    and substantially lower than MPFR.
  \name is more accurate than QD
    and substantially more accurate than \eftsan,
    placing it at a particularly practical point
    in the speed-accuracy spectrum.
  }
  \label{fig:sci-runtime}
\end{figure}

The results above demonstrate
  that \name significantly improves on \eftsan,
  with further improvements on challenging numerical benchmarks
  enabled by \RO.
However, although \eftsan is the current state of the art
  among floating-point debuggers that use
  machine-float residues,
  other debuggers instead use
  higher-precision residues.
We therefore compare \name, with \RO enabled,
  to both the ground-truth MPFR-based debugger
  and a debugger based on the QD library,
  which computes residues from quad-double shadow values.

Among the three debuggers,
  MPFR is the most accurate by definition,
  since we use it to obtain the ground-truth results.
The comparison between \name and QD is therefore more interesting.
Across all evaluated benchmarks, \name and QD differ on \numQDInteresting~benchmarks;
  \name achieves better results on \numQDImproved~of them (\pctQDImproved),
  \numQDMixed~exhibit a mixed pattern of improvements and regressions
  (one debugger produces more false positives while the other produces more false negatives),
  while QD achieves better results on just \numQDLoss.
In other words, lower-precision residues combined with \RO
  provide better accuracy, on average,
  than higher-precision residues without \RO.

Moreover, \name is also faster than the QD or MPFR debuggers.
We instrumented each debugger to measure
  only the execution time of floating-point operations
  (including all shadow operations or residue computations)
  using an \texttt{rdtsc}-based cycle counter.
\Cref{fig:sci-runtime} shows the runtime overhead,
  compared to uninstrumented execution,
  across all scientific benchmark suites
  for the \eftsan, \name, QD, and MPFR debuggers.
Note that the horizontal axis is logarithmic.
All debuggers are slower than uninstrumented execution:
  \eftsan is the fastest, \name is somewhat slower,
  while QD and MPFR are slower still.
This overhead arises from the cost of performing each floating-point operation
  using software-based high-precision arithmetic.
Although \name \emph{is} slower than \eftsan,
  due to the additional operation tracking required for \RO,
  it is substantially more accurate,
  placing \name at a favorable balance
  between high accuracy and low overhead.
\section{Related Work}

Numerical analysis and the correctness of numerical software
  have long been studied problems in computer science~\cite{mathematics-sand,higham-book}.
More recently, researchers have developed automated tools
  for deriving error bounds or finding bugs in numerical programs.

\paragraph{Error Bound Tools}
Some of the earliest analysis tools for numerical software
  used abstract interpretation to bound rounding error.
For example, Salsa~\cite{salsa} represented program states
  using pairs of intervals---%
  one for variable values and one for errors.
Conceptually, this resembles a residue-based technique,
  using interval arithmetic to bound residues.
Later tools followed the same separation between values and errors
  but used different methods to bound the error intervals.
For example, Rosa~\cite{rosa} and Daisy~\cite{daisy}
  used affine arithmetic to obtain tighter error bounds.
The FPTaylor project~\cite{fptaylor}, likewise,
  modeled values and errors separately using an
  ``error Taylor series''---%
  a generalization of affine arithmetic---%
  and then used a nonlinear global optimizer
  to bound the maximum error.
Later tools such as Satire~\cite{satire} follow a similar approach.
\name uses a similar residue-based idea,
  though it performs its analysis dynamically at runtime.
In contrast, error-bound tools perform their analysis symbolically,
  compute ranges rather than specific error values,
  and optimize for tight error bounds rather than runtime overhead.

More recently, the NumFuzz approach~\cite{numfuzz}
  has explored faster numerical analysis
  using techniques inspired by type checking,
  including extensions to backward error analysis~\cite{bean}.
This type-theoretic approach differs substantially
  from the value/error separation used by previous tools,
  focusing on error sensitivity rather than the exact range of values or errors.
However, this technique has not yet been extended
  to handle subtraction and cancellation,
  let alone complex scientific software.

One challenge for many numerical analysis techniques
  is control-flow divergence between
  the floating-point execution and the ideal real execution.
For example, Seesaw~\cite{seesaw}
  analyzes conditional expressions to determine
  whether numerical error could cause a conditional
  to evaluate differently.
This problem is especially difficult for analysis tools,
  which must consider a range of inputs.
Floating-point debuggers like \name, by contrast,
  typically follow the floating-point control flow
  while reporting potential points of divergence.

\paragraph{Floating-point Debugging}
In contrast to analysis tools,
  floating-point debuggers execute the program
  on specific inputs
  and identify cases where rounding error
  accumulates enough to significantly deviate
  from the ideal real-number result.
Because they analyze a specific execution,
  floating-point debuggers can often be more precise:
  they can explicitly detect control-flow divergence
  and compute exact error values for intermediate variables.

Early debuggers such as FpDebug~\cite{fpdebug}
  used arbitrary-precision shadow values
  to detect operations like subtraction that cause cancellation.
This idea was developed further in Herbgrind~\cite{herbgrind},
  which introduced the concept of \emph{spots}---%
  program outputs or control-flow decisions
  where floating-point error produces user-visible effects.
Herbgrind then applies anti-unification
  to identify a minimal floating-point expression tree
  responsible for these effects.
However, these early debuggers were relatively slow
  because they relied on arbitrary-precision arithmetic.

Later work focused on improving performance.
FPSanitizer~\cite{fpsanitizer}
  integrated shadow computations and error reporting
  directly into the compilation pipeline as an LLVM pass
  and also explored multi-threaded debugging
  to reduce overhead~\cite{pfpsanitizer}.
However, arbitrary-precision arithmetic remained a bottleneck.
EFTSanitizer addressed this by replacing
  arbitrary-precision shadow values
  with residues computed directly using machine floating-point operations
  and error-free transformations.
Combined with FPSanitizer's optimized compilation process,
  EFTSanitizer achieved overheads of about $10\times$
  while also computing dependency graphs
  to help developers understand error propagation.
Similarly, Shaman~\cite{shaman} uses error-free transformations
  to identify values with high numerical error,
  though it requires users to instrument specific parts of the program.
However, abandoning arbitrary precision
  leads to higher rates of false positives and false negatives,
  a gap that \name aims to close.

Meanwhile, several tools avoid shadow-value computation entirely.
For example, ATOMU~\cite{atomu}
  computes condition numbers for each floating-point operation
  and issues warnings when the condition number is large.
Because ATOMU does not track shadow values,
  it avoids the question of how precise those values must be.
Explanifloat~\cite{explanifloat} extends this idea further,
  combining error-free transformations,
  a logarithmic number system,
  and condition numbers
  to detect subtle rounding errors.
However, condition-number-based approaches
  tend to produce many false reports.
Explanifloat, for example, reports a precision of about 80\%,
  meaning that roughly 20\% of its warnings are false.

The BZ~\cite{baozhang} and RAIVE~\cite{raive} tools
  also avoid storing shadow values,
  instead inspecting the exponents of computed floating-point values
  to detect potential numerical issues.
These approaches likewise suffer from high false-report rates.
Finally, the Verrou project~\cite{verrou}
  uses stochastic rounding---%
  effectively perturbing floating-point values---%
  to study the impact of rounding error.
Like \name, this approach relies on repeated executions
  to understand numerical error.
However, \name deterministically identifies
  which residues interfere with one another,
  while Verrou is fully stochastic
  and therefore requires many more executions
  to accurately estimate program error.
\section{Conclusion}

Floating-point debuggers can help programmers identify and,
  ultimately, fix numerical issues.
However, even the best existing debuggers have limited accuracy,
  which can make their results misleading.
In this work, we improved residue-based debugging
  at two levels.
First, we refined the machinery for computing machine-float residues:
  how rounding errors are estimated
  and how those errors are propagated through residue functions.
These changes made the debugger's shadow computation
  more faithful to the underlying real-number execution,
  reducing false reports in
  \numSciBenchmarksImproved~out of \numSciInterestingBenchmarks~scientific workloads
  where prior tools produce misleading results,
  without sacrificing the efficiency of machine-precision shadow values.

We then showed that some remaining failures
  were not merely due to bugs or approximations in residue formulas,
  but stemmed from a more fundamental limitation of fixed-precision residues.
Under absorption,
  a residue could preserve the dominant contribution needed at one point
  in the computation
  or the smaller contribution needed after a later cancellation,
  but not both within a single execution.
The \RO framework addressed this limitation
  by separating these measurements across multiple runs:
  silenced runs revealed information that ordinary execution hides,
  and the final override run reintroduced corrected residues
  at the points where they are needed.
The idea was further extended to handle
  more complex cases that arise in practice,
  including repeated silencing
  and multiple interacting cancellations.
On \numMFInteresting~benchmarks drawn from numerical analysis papers
  and textbooks where initial runs failed to achieve perfect results,
  \RO reduced false report counts in \numROTotalImproved~of them
  while requiring only a modest number of re-executions on average.
Overall, these results show that substantially more accurate
  floating-point debugging does not require
  high-precision shadow computation.
Instead, improved residue computation,
  combined with targeted override runs for absorption,
  recovers much of that accuracy at far lower cost.

\section{Data Availability Statement}
We will submit the \name debugger,
  its driver program, all benchmarks,
  and the evaluation setup for Artifact Evaluation.
Upon submission, we will make the \name repository public,
  containing all of the above components.
\name will be released under the MIT license
  (the same license as the underlying \texttt{wasm3} interpreter),
  while the benchmarks will be released
  under their respective licenses.
No proprietary data or closed-source components are required
  to reproduce our results.
The full evaluation takes a few hours to run,
  and we expect artifact evaluators to reproduce it in full.

\bibliographystyle{ACM-Reference-Format}
\bibliography{references}

\end{document}